\documentclass[seceq]{ptptex}

\usepackage{graphicx}
\title{%        %You can use \\ for explicit line-break
Thermal Width Broadening of the $0^{++}$ Glueball Spectrum near $T_c$
}

\author{%       %Use \scshape  for the family name
Noriyoshi \textsc{Ishii}$^1$, Hideo \textsc{Suganuma}$^2$,
and
Hideo \textsc{Matsufuru}$^3$
}

\inst{%         %Affiliation, neglected when [addenda] or [errata]
$^1$Radiation Laboratory,
RIKEN (The Institute of Physical and Chemical Research),
2--1 Hirosawa, Wako, Saitama 351--0198, Japan\\
$^2$Faculty of Science, Tokyo Institute of Technology,
2--12--1 Ohkayama, Meguro, Tokyo 152--8552, Japan\\
$^3$Yukawa Institute for Theoretical Physics, Kyoto University,
Kitashirakawa-Oiwake, Sakyo, Kyoto 606--8502, Japan
}

\abst{%         %this abstract is neglected when [addenda] or [errata]
We  study  the $0^{++}$  glueball  correlator  constructed with  SU(3)
anisotropic quenched  lattice QCD  at various temperature  taking into
account  the   possible  existence  of   the  thermal  width   in  the
ground-state  peak.   For  this  purpose, we  adopt  the  Breit-Wigner
ansatz, analysing  the lattice  data obtained with  5,500--9,900 gauge
configurations  at  each $T$.  The  results  indicate the  significant
thermal  width  broadening  as  $\Gamma(T_c)  \sim  300$  MeV  with  a
reduction  in the peak  center as  $\Delta\omega_0(T_c) \sim  100$ MeV
in the vicinity of the critical temperature $T_c$.  }

\newcommand{\Eq}[1]{Eq.~({\ref{#1}})}
\newcommand{\Fig}[1]{Fig.\protect\ref{#1}}

\newcommand{\Tate}{\rule{0cm}{1.1em}}
\newlength{\Tatescale} 

\newlength{\figwidth} 
\begin{document}

\maketitle

\section{Introduction}
At finite  temperature/density, QCD changes its  vacuum structure such
as  the   reduction  of  the   string  tension,  the   partial  chiral
restoration, etc, even below the critical temperature $T_c$ of the QCD
phase transition.
These  changes are  expected  to  affect the  hadrons  leading to  the
changes  in  the  various   hadronic  properties,  since  hadrons  are
composite particles consisting of quarks and gluons.
The  hadronic pole-mass  shifts are  thus considered  to serve  as the
important precritical  phenomena of the QCD phase  transition near the
critical temperature $T_c$, and have been extensively studied by using
various      QCD-motivated      low-energy     effective      theories
\cite{hatsuda,miyamura,hatsuda2,ichie}.   These  studies  suggest  the
pole-mass reductions  of charmoniums,  light $q\bar{q}$ mesons  and the
glueball near the critical temperature.

As for the lattice QCD, it  has been more popular to study the spatial
correlations (the screening mass)  than the temporal correlations (the
pole-mass) at finite temperature.
This is  because the temporal extension  of the lattice  is subject to
$1/T$.  Accordingly, number of  the temporal lattice data decreases at
finite temperature, which used to be the main obstacle for the lattice
QCD to measure the hadronic pole-mass.
Recently, by using the anisotropic  lattice, which has a finer lattice
spacing in the  temporal direction than in the  spatial direction, the
accurate measurement of  the pole-mass with the lattice  QCD has become
possible  at finite  temperature  \cite{klassen,taro}.  Quenched-level
Monte  Carlo  calculations  reveal   the  profound  results  that  the
pole-masses of  the $q\bar{q}$ mesons  are almost unchanged  from their
zero-temperature  values in  the confinement  phase \cite{taro,umeda},
while  the pole-mass  of  the  $0^{++}$ glueball  shows  the $300$  MeV
reduction near the critical temperature \cite{ishii}.
Although  these thermal  effects are  weaker than  anticipated  by the
effective  model studies,  these  tendencies are  consistent with  the
recent quenched-level  lattice studies on the screening  mass at finite
temperature \cite{laermann,gupta}.
In these analysis, the bound-state  peaks in the spectral function are
assumed  to be narrow  enough.  However,  at finite  temperature, each
bound-state acquires  the thermal  width through the  interaction with
the  heat  bath.  The  thermal  width  is expected  to  grow  up  with
temperature, which may lead to  a possible collapse of the narrow-peak
assumption in some cases.  In this paper, we first discuss what is the
expected  consequence  of  the  thermal width  broadening.   Then,  we
propose the Breit-Wigner ansatz  for the fit-function for the temporal
correlator at finite  temperature. We finally show the  results of the
Breit-Wigner analysis  of the  $0^{++}$ glueball correlator  at finite
temperature \cite{ishii2}.
\section{The Breit-Wigner Ansatz}
We  begin  by  considering  the temporal  correlator  $G(\tau)  \equiv
Z(\beta)^{-1}\mbox{Tr}\left(  e^{-\beta  H}  \phi(\tau)\phi(0)\right)$
and its spectral representation as
\begin{eqnarray}
	G(\tau)
=
	\int_{-\infty}^{\infty}
	{d\omega \over 2\pi}
	{
		\cosh\left(\omega(\beta/2 - \tau)\right)
	\over
		2\sinh(\beta\omega/2)
	}
	\rho(\omega),
\label{spectral.representation}
\end{eqnarray}
where    $H$   denotes    the    QCD   Hamiltonian,    $Z(\beta)\equiv
\mbox{Tr}(e^{-\beta H})$  the partition function,  $\phi(\tau)$ is the
zero-momentum projected glueball  operator \cite{ishii} represented in
the  imaginary-time Heisenberg picture  as $\phi(\tau)  \equiv e^{\tau
H}\phi(0)e^{-\tau  H}$.   Here,  $\rho(\omega)$ denotes  the  spectral
function
\begin{eqnarray}
	\rho(\omega)
&\equiv&
	\sum_{m,n}
	{ |\langle n|\phi|m \rangle |^2 \over Z(\beta) }
	e^{-\beta E_m}
	\times 2\pi
	\left(\Tate
		\delta(\omega - \Delta E_{nm})
	-	\delta(\omega + \Delta E_{mn})
	\right),
\label{spectral.function}
\end{eqnarray}
where $E_n$  denotes the energy  of $n$th excited states,  and $\Delta
E_{mn} \equiv E_m - E_n$.  Note that $\rho(\omega)$ is odd in $\omega$
reflecting  the  bosonic nature  of  the  glueball.   By adopting  the
appropriate ansatz for $\rho(\omega)$, we can extract various physical
quantities such  as the pole-mass  and the width through  the spectral
representation \Eq{spectral.representation}.

We  first consider  the  case  where the  bound-state  peak is  narrow
\cite{taro,umeda,ishii}.    In   this   case,   by   introducing   the
temperature-dependent   pole-mass   $m(T)$,   $\rho(\omega)$   can   be
parameterized as
\begin{equation}
	\rho(\omega)
\simeq
	2\pi A
	\left(\Tate
		\delta(\omega - m(T))
	-	\delta(\omega + m(T))
	\right),
\label{single.pole}
\end{equation}
where  $A$  represents the  strength.   The  second delta-function  is
introduced  to  respect  the  odd-function nature  of  $\rho(\omega)$.
Since  the  corresponding $G(\tau)$  reduces  to  a single  hyperbolic
cosine,  the  pole-mass  measurement  at  finite  temperature  can  be
performed in  the same  way as the  standard mass measurement  at zero
temperature \cite{morningstar}.

We next  consider the case where  the thermal width is  wide.  In this
case,  the peak  center  $\omega_0$ of  $\rho(\omega)$ represents  the
observed ``mass'' of the thermal hadron.  What follows the narrow-peak
assumption in this case ?   To consider this, we notice that $G(\tau)$
can be thought of as a weighted average of hyperbolic cosines with the
weight as
\begin{equation}
	W(\omega)
\equiv
	\frac{
		\rho(\omega)
	}{
		2\sinh(\beta\omega/2)
	}.
\end{equation}
Here,  $2\sinh(\beta\omega/2)$ in  the denominator  works as  a biased
factor, which  enhances the smaller $\omega$  region while suppressing
the larger $\omega$ region.
Consequently,  the pole-mass  $m(T)$, which  is approximated  with the
peak  position  of  $W(\omega)$,  is  smaller  than  the  peak  center
$\omega_0$ of the spectral function $\rho(\omega)$, i.e., the observed
hadron ``mass'' \cite{ishii2}.
What is the appropriate functional form of the fit-function ?  To find
this,  we consider the  retarded Green  function $G_{\rm{R}}(\omega)$.
At $T=0$, bound-state poles of $G_{\rm{R}}(\omega)$ are located on the
real $\omega$-axis.  At $T > 0$, bound-state poles are moving into the
complex  $\omega$-plane with increasing  temperature.  Suppose  that a
bound-state pole is located at $\omega = \omega_0 - i\Gamma$ as
\begin{equation}
	G_{\rm{R}}(\omega)
=
	{A \over \omega - \omega_0 + i\Gamma} + \cdots,
\end{equation}
where $A$  represents the  residue at the  pole, and  ``$\cdots$'' the
non-singular terms  around the pole.   Since the spectral  function is
the imaginary part of the retarded Green function, the contribution of
this complex pole is expressed in the form of Lorentzian as
\begin{eqnarray}
	\rho(\omega)
&=&
	- 2 \mbox{Im}\left(\Tate G_{\rm{R}}(\omega)\right).
\label{single.peak}
\\\nonumber
&\simeq&	
	2\pi A
	\left(\Tate
		\delta_{\Gamma}(\omega - \omega_0)
	-	\delta_{\Gamma}(\omega + \omega_0)
	\right),
\end{eqnarray}
where     $\displaystyle    \delta_{\epsilon}(x)     \equiv    {1\over
\pi}\mbox{Im}\left(  {1\over x -  i\epsilon} \right)  = {1  \over \pi}
{\epsilon \over  x^2 + \epsilon^2}$  is a smeared  delta-function with
the  width $\epsilon  > 0$.  The  second term  in \Eq{single.peak}  is
introduced to respect the odd-function nature of $\rho(\omega)$.
%%%%%%%%%%%%%%%%%%%%%%%%%%%%%%%%%%%%%%%%%%%%%%%%%%%%%%%%%%%%%%%%%%%%%%%%%%%%%%%
In   the   limit   $\Gamma\to   +0$,   \Eq{single.peak}   reduces   to
\Eq{single.pole}.
%%%%%%%%%%%%%%%%%%%%%%%%%%%%%%%%%%%%%%%%%%%%%%%%%%%%%%%%%%%%%%%%%%%%%%%%%%%%%%%
We  thus  see that  the  appropriate  fit-function,  which takes  into
account the  effect of  the non-zero thermal  width, is  the following
Breit-Wigner type as
\begin{equation}
	g(\tau)
\equiv
	\int_{-\infty}^{\infty}
		{d\omega \over 2\pi}
		{
			\cosh\left(\omega(\beta/2 - \tau)\right)
		\over
			2\sinh(\beta\omega/2)
		}
	\times
		2\pi \widetilde A
		\left( \Tate
			\delta_{\Gamma}(\omega - \omega_0)
		-	\delta_{\Gamma}(\omega + \omega_0)
		\right),	
\label{breit-wigner}
\end{equation}
where  $\widetilde  A$,  $\Gamma$  and $\omega_0$  are  understood  as
fit-parameters, corresponding  to the  residue, the thermal  width and
the peak center, respectively.
%%%%%%%%%%%%%%%%%%%%%%%%%%%%%%%%%%%%%%%%%%%%%%%%%%%%%%%%%%%%%%%%%%%%%%%%%%%%%%%
Note  that  $g(\tau)$  is  a  generalization of  the  ordinary  single
hyperbolic cosine fit-function.
%%%%%%%%%%%%%%%%%%%%%%%%%%%%%%%%%%%%%%%%%%%%%%%%%%%%%%%%%%%%%%%%%%%%%%%%%%%%%%%
In order to use Eqs.(\ref{single.pole}) and (\ref{single.peak}), it is
essential  to suppress  the  higher spectral  contributions.  This  is
usually achieved by appropriate choices  of the fit-ranges and also by
improving  the  glueball operator,  for  instance,  with the  smearing
method \cite{ishii2}.
\section{Numerical Result}
We   use   the  SU(3)   anisotropic   lattice   plaquette  action   as
\begin{equation}
	S_{\rm{G}}
=
	{\beta_{\rm{lat}}\over N_c}{1\over\gamma_{\rm{G}}}
	\sum_{s,i<j\le   3}
	\mbox{Re}\mbox{Tr}( 1  - P_{ij}(s) )
+
	{\beta_{\rm{lat}}\over    N_c}\gamma_{\rm{G}}
	\sum_{i,j\le  3}
	\mbox{Re}\mbox{Tr}( 1 -  P_{i4}(s) ),
\end{equation}
where $P_{\mu\nu}(s) \in  {\rm{SU(3)}}$ denotes the plaquette operator
in  the  $\mu$-$\nu$-plane.   The   lattice  parameter  and  the  bare
anisotropic parameter are  fixed as $\beta_{\rm{lat}}\equiv 2N_c/g^2 =
6.25$ and $\gamma_{\rm{G}} = 3.2552$, respectively, so as to reproduce
renormalized anisotropy $\xi  \equiv a_s/a_t = 4$ \cite{klassen,taro}.
These  parameter  set  reproduces   $a_s^{-1}  =  2.341(16)$  GeV  and
$a_t^{-1} =  9.365(66)$ GeV, where  the scale unit is  introduced from
the on-axis data  of the static inter-quark potential  with the string
tension  $\sqrt{\sigma}  =   440$  MeV.   Numerical  calculations  are
performed on  the lattice of the  size $20^3 \times  N_t$ with various
$N_t$.  The  critical temperature $T_c$  on this lattice  is estimated
from the  behavior of the Polyakov-loop susceptibility  as $T_c \simeq
280$ MeV. The pseudo-heat-bath algorithm  is adopted for the update of
the gauge configurations.  In order to construct the temporal glueball
correlators,  we  use  5,500   to  9,900  gauge  configurations.   The
statistical data are  divided into bins of the size  100 to reduce the
possible  auto-correlations   near  the  critical   temperature.   The
smearing  method is  used to  obtain the  improved  glueball operator,
which  is   determined  by  examining  its  behavior   at  the  lowest
temperature, i.e., $T=130$ MeV.

\begin{figure}
\begin{center}
\includegraphics[width=0.7\figwidth,angle=-90]{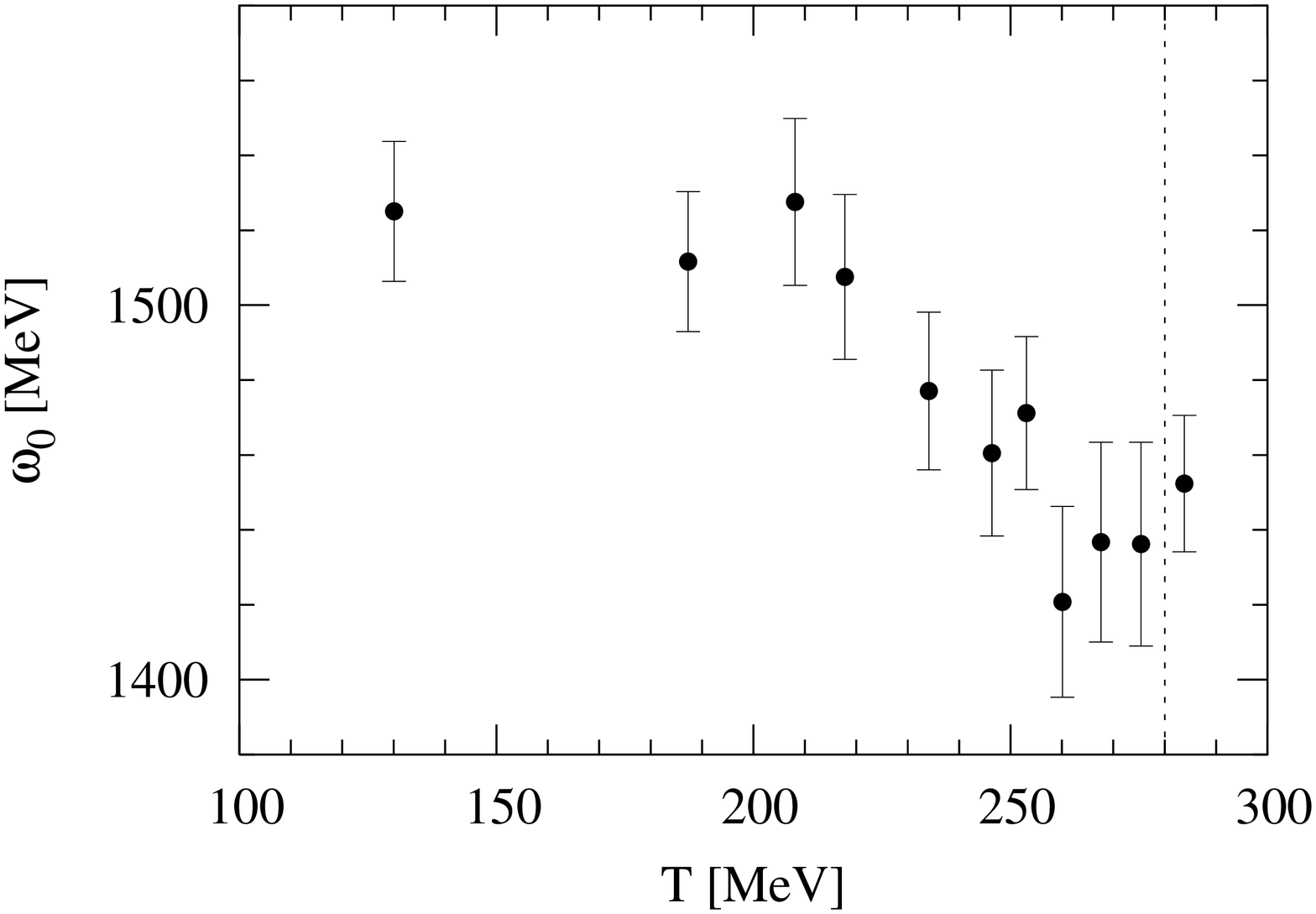}
\includegraphics[width=0.7\figwidth,angle=-90]{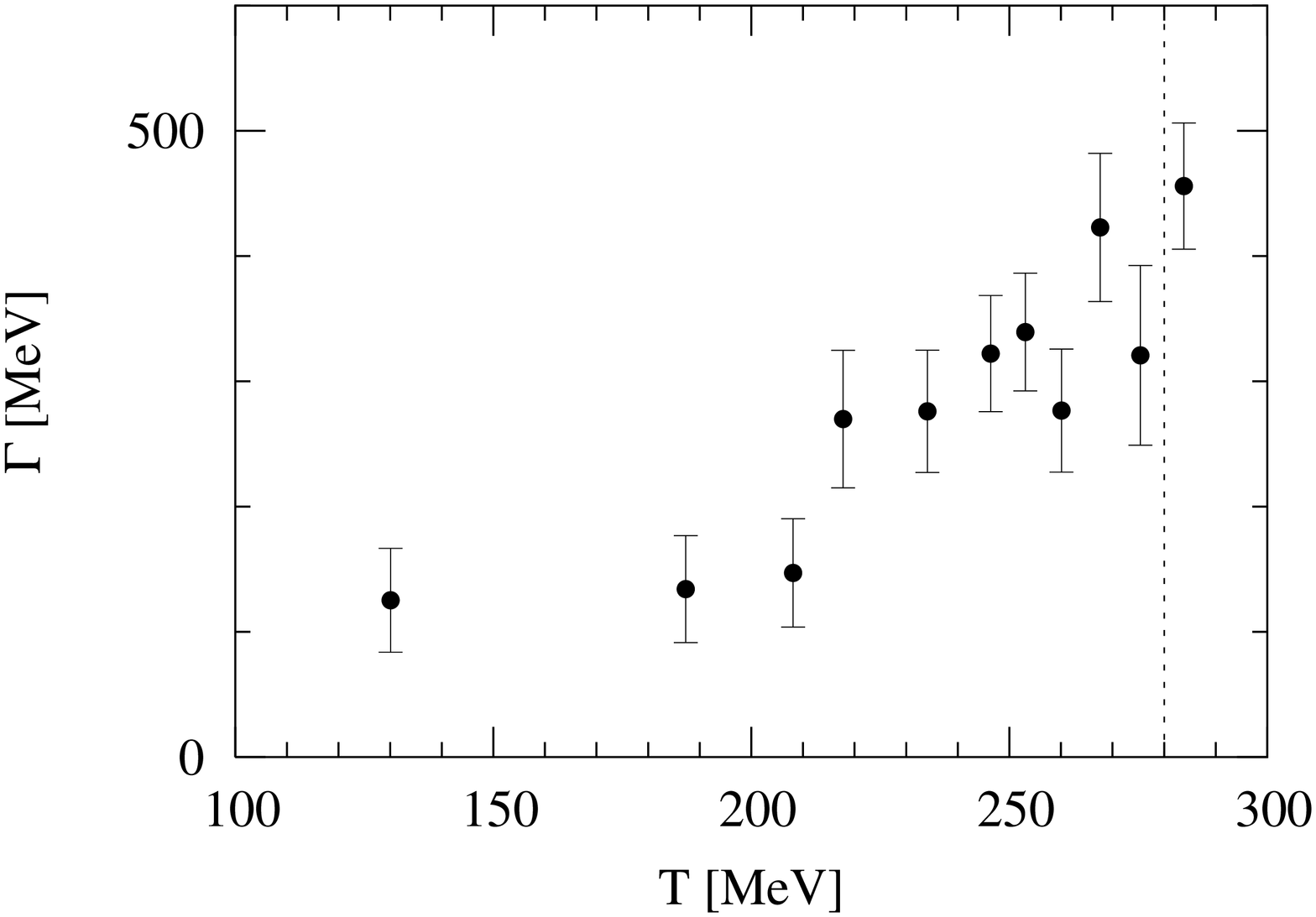}
\end{center}
\caption{The center  $\omega_0$ and the thermal width  $\Gamma$ of the
lowest  $0^{++}$ glueball  peak plotted  against temperature  $T$. The
vertical  dotted lines  indicate the  critical  temperature $T_c\simeq
280$ MeV.}
\label{figure}
\end{figure}
The peak center and the thermal width of thermal glueball are obtained
by the  best-fit analysis of the suitably  smeared glueball correlator
$G(\tau)/G(0)$  with \Eq{breit-wigner}  at  various temperatures.   In
\Fig{figure},  the  peak  center  $\omega_0$  and  the  thermal  width
$\Gamma$  are plotted against  temperature.  While  narrow-peak ansatz
leads  to   the  pole-mass   reduction  of  300   MeV  near   $T_c$  in
Ref.\cite{ishii},   the  Breit-Wigner   analysis  indicates   a  small
reduction in the  peak center as $\Delta \omega_0(T_c)  \sim 100$ MeV.
Instead,  we  observe  a   significant  thermal  width  broadening  as
$\Gamma(T_c) \sim 300$ MeV.
%%%%%%%%%%%%%%%%%%%%%%%%%%%%%%%%%%%%%%%%%%%%%%%%%%%%%%%%%%%%%%%%%%%%%%
\section{Summary and Discussion}
We have  studied the temporal  correlator of the $0^{++}$  glueball at
finite  temperature using  SU(3) anisotropic  lattice QCD  at quenched
level with 5,500 to 9,900 gauge configurations at each temperature.
We have proposed the Breit-Wigner  ansatz for the fit-function to take
into  account the  effect  of  the non-zero  thermal  width at  finite
temperature.
We  have applied the  Breit-Wigner analysis  to the  temporal glueball
correlator  at finite  temperature,  and have  observed a  significant
broadening of the  thermal width as $\Gamma(T_c) \sim  300$ MeV with a
slight reduction of the peak center as $\Delta \omega_0(T_c) \sim 100$
MeV in the vicinity of the critical temperature $T_c$.

In our present analysis, the  Breit-Wigner ansatz seems to be the most
appropriate one.
Above  $T_c$, however,  the particle  spectrum is  expected  to change
drastically, and a proper  assumption on the spectral function becomes
less trivial.
In this  respect, it  is interesting to  apply the  recently developed
maximum  entropy method  (MEM) \cite{mem,asakawa},  which  requires no
assumption on the shape of the spectral function.

Comparing our results\cite{ishii,ishii2}  with the related lattice QCD
results \cite{taro,umeda}, the quenched  lattice QCD indicates that no
significant thermal effects on the pole-mass have been so far observed
in  the  $q\bar{q}$  sector\cite{taro,umeda},  whereas  a  significant
thermal  effects  can  be  observed  in  the  glueball  sector.  These
behaviors  are consistent with  the thermal  effects on  the screening
mass \cite{laermann,gupta}.

To discuss the experimental observability, it is important to estimate
the  effects of  the dynamical  quarks.  One  of the  most significant
contributions of the  light dynamical quarks in unquenched  QCD is the
``string  breaking''  phenomenon,  i.e.,  the  light  dynamical  quark
screens the  colored flux and the  long flux tubes are  expected to be
broken into pieces. The string breaking thus affects the long distance
behavior  of the inter-particle  interactions, which  could lead  to a
possible changes in  the structures of hadrons in  quenched QCD. In the
case of  the $0^{++}$ glueball, since  its size is known  to be rather
compact as  $R\simeq 0.4$ fm \cite{ishii,ishii2}, such  changes of the
glueball structure is expected to be less significant.
As  for the  mixing  effect  with the  $0^{++}$  $q\bar{q}$ mesons  in
unquenched QCD, a recent unquenched lattice QCD study shows that there
is no quark mass dependence  on the glueball mass, which suggests that
the mixing is actually rather small.
However, it  is in  principle desirable to  estimate the size  of this
mixing explicitly, which is left for the future studies.

%
%% In  the $q\bar{q}$  sector (both  light and  heavy  $q\bar{q}$), these
%% quenched lattice QCD results are inconsistent with the effective model
%% studies \cite{hatsuda,hatsuda2,miyamura}.
%% %
%% In  the  case of  the  light $q\bar{q}$,  the  main  driving force  is
%% expected  to be  the  partial restoration  of  the spontaneous  chiral
%% symmetry  breaking, where  a significant  contribution from  the light
%% dynamical  quarks  are  expected.  Hence,  the  inconsistency  may  be
%% resolved by unquenched calculations.
%% %
%% In contract to this, in the case of the heavy $q\bar{q}$, the quenched
%% approximation  makes  the   correspondence  between  the  lattice  QCD
%% calculation  and  the effective  model  calculation (potential  model)
%% better, and it is difficult  to find the origin of this inconsistency.
%% Inclusion of  dynamical quarks could  not improve the  situation.  The
%% expected  changes due  to the  inclusion of  heavy dynamical  quark is
%% small. The  light dynamical quarks may  contribute to a  change of the
%% charmonium  spectra,  which,  however,  does  not  correspond  to  the
%% potential model  calculations. Something essential must  be missing in
%% this system, which are left as an open problem for future studies.

\section*{Acknowledgments}
The lattice QCD Monte Carlo calculations have been performed partly on
NEC-SX5 at Osaka University and partly on HITACHI SR8000 at KEK.

%\section*{Acknowledgements}
%We would like to thank ...........

%\appendix
%\section{First Appendix} %Empty argument \section{} yields `Appendix'. 
%
%\section{Second Appendix}

\end{document}